\documentstyle[prl,aps,amsfonts,amssymb,twocolumn,epsfig]{revtex} 

\preprint{TH97/1} \title{ Field-induced quasi-periodic coherence 
effects in certain small magnets.  } \author{ S. E. Barnes }

\address{ Department of Physics, University of Miami, Coral Gables, 
Florida 33124 }

\begin{document}
\draft

\twocolumn[
\hsize\textwidth\columnwidth\hsize\csname @twocolumnfalse\endcsname

\date{\today} \maketitle
\begin{abstract}
Small ferromagnets and anti-ferromagnets with an easy-plane anisotropy 
have a ground to first excited state (tunnel) splitting which is 
quasi-periodic in the magnitude of a field applied perpendicular to a 
principal anisotropy axis.  The associated oscillations in 
thermodynamic quantities might be used to prove the existence of a 
coherent ground state even when the tunnel splitting itself cannot be 
directly detected.
\end{abstract}

\vskip1.0pc]

The study of small magnets tests our understanding of the transition 
between quantum and classical physics.  In the classical regime the 
order parameter, i.e., the magnetization, of a ferromagnet points is 
one of a number of equivalent directions which correspond to a minimum 
of the macroscopic anisotropy energy.  On the other hand in the 
quantum world of small magnets the ground state comprises a coherent 
superposition of all of these equivalent equilibrium directions.  
While there is now fairly strong evidence\cite{thomas} of {\it 
relaxation\/} via quantum {\it tunneling\/} in Mn${}_{12}$ the 
observation of a {\it coherent ground state\/} remains an elusive 
goal.  The only experimental evidence, for the anti-ferromagnet 
ferritin\cite{awschalom}, remains highly controversial.

The principal purpose of this Letter is to show, for small 
ferromagnets and anti-ferromagnets with an {\it easy-plane\/} 
anisotropy, there exist rather dramatic level crossing effects.  
Unlike similar easy-axis magnets which have tunneling effects very 
sharply peaked at level crossings, the tunnel splitting {\it 
oscillates\/} as a function of the magnet field.  The period, e.g., 
$\Delta H = 2 H_{\parallel}$ for a ferromagnet, is determined by an 
easy-to-measure bulk anisotropy field $H_{\parallel}$.  The 
observation of this oscillatory signature can provide an unrefutable 
proof of coherence, and since $M = - \partial F / \partial H$ where 
$F$ is the free energy, there is necessarily an oscillatory component 
in the magnetization, and other thermodynamic quantities, which might 
be readily detected even if the splitting cannot.

The ferromagnet is modeled by a single large spin subject to the 
external and anisotropy fields\cite{chud}, i.e., the Hamiltonian 
${\cal H} = g \mu_{B}\vec S \cdot \vec H + K_{\parallel} {S_{z}}^{2} + 
K_{\perp}{S_{x}}^{2}$ where without loss of generality it is assumed 
that $|K_{\parallel}|>|K_{\perp}|$.  For an easy plane magnet 
$K_{\parallel} > 0$.  The energy parameters $ K_{\parallel}$, 
$K_{\perp}$ and, for the anti-ferromagnet, $J$ all scale as $S^{-1}$.  
The equivalent physical quantities are $H_{\parallel} = 
K_{\parallel}/S$, $H_{\perp} = K_{\perp}/S$ and $H_{e} = J/S$.  This 
is important in ratios such as $(h/K_{\parallel}) =(S h 
/H_{\parallel})$.

The problem is formulated\cite{seb,sebetal} in terms of auxiliary 
particles.  Consider a ferromagnet.  A basis $| S_z> \equiv |n>$ is 
chosen.  Then an auxiliary particle, a fermion $f^{}_n$, is associated 
with each state via the mapping $|n> \to f^\dagger_n |>$ where $|>$ is 
a non-physical vacuum without any auxiliary particles.  Defined is a 
bi-quadratic version of an operator $\hat O$ via: $ \hat O \to 
\sum_{n,n^\prime} f^\dagger_n <n|\hat O|n^\prime> f^{}_{n^\prime}$.  
The constraint $ Q = \sum_n \hat n_n = \sum_n f^\dagger_n f^{}_n =1 $.  
It has been shown \cite{seb} that such schemes preserve all operator 
multiplication rules including commutation rules. The replacement rule is 
applied to the Hamiltonian ${\cal H}$ to yield, taking $g 
\mu_{B} \vec H = - h \hat z$:
\begin{eqnarray}
&{\cal H} = \sum_n \Big( K_{\parallel} {n}^2- n h + \nonumber\\
&{1\over 4} 
K_{\perp}[ \left( M_{n}^{n+1}\right)^{2} + 
\left(M_{n}^{n-1}\right)^{2}] \Big) f^\dagger_n f^{}_n\nonumber\\
&+ 
{1\over 4} K_{\perp} \sum_n M_{n}^{n+1} M_{n+1}^{n+2} ( 
f^\dagger_{n+2} f^{}_n + H.c.  ){,}
\label{une}
\end{eqnarray}
where the $M_{n}^{n+1} =[S(S+1) - n(n+1)]^{1/2}$ are the matrix 
elements of $S^{\pm}$.  This is {\it two\/} tight binding models of 
spinless fermions $f^\dagger_n$.  The constraint $Q=1$ implies this is 
a single particle problem.

\begin{figure}[t]
\epsfig{width=0.95 \linewidth,file=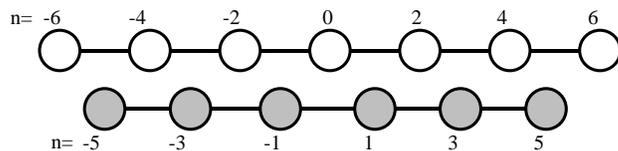} \caption{ 
The two chains for integer spin each have symmetry about $n=0$}
\label{A}
\end{figure}

\begin{figure}[t]
\epsfig{width=0.95 \linewidth,file=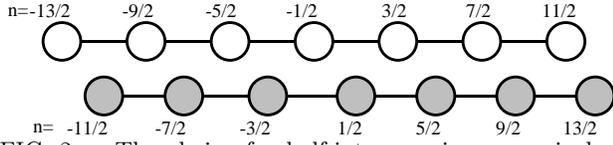} \caption{ 
The chains for half-integer spin are equivalent though the map  $n 
\leftrightarrow -n$}
\label{B}
\end{figure}

The two ``chain'' structure reflects the fact that the ``hopping'' 
term in Eqn.  (\ref{une}) couples ``sites'' with indices which differ 
by two.  This structure implies immediately a spin-parity effect found 
by Loss et al.\cite{loss}, and von Delft and Henley\cite{henley}.  For 
the case of integer spin, Fig.  \ref{A}, the two chains comprise the 
sites $n = -S,-(S-2), -(S-3) \ldots (S-3), (S-2), S$ and $n = -(S-1), 
-(S-3), -(S-5), \ldots (S-5), (S-3), (S-1)$ which are both symmetric 
relative to $n=0$.  On the other hand for half-integer spin, Fig.  
\ref{B}, the chains are $n = -S,-(S-2), -(S-3) \ldots (S-5), (S-3), 
(S-1)$ and $n = -(S-1), -(S-3), -(S-5), \ldots (S-3), (S-2), S$, which 
are equivalent to each other through the map $n \to - n$ but which 
lack the symmetry about $n=0$.  Because of the equivalence of the two 
chains there must always be a double (Kramers') degenerate ground 
state, without a tunnel splitting, for half-integer spin, while 
because of symmetry about $n=0$, there can be tunnel split pairs for 
integer spin.

There is a key {\it fixed point\/}\cite{seeseb} when $K_{\perp} \to 
0$.  All off-diagonal matrix elements are zero, the diagonal energies 
are $K_{\parallel} {n}^2 - hn$, and for an {\it integer spin 
easy-plane ferromagnet\/}, i.e., with $K_{\parallel} >0$, and in zero 
field ($h=0$), the ground state lies on the site $n=0$, and therefore 
the even site chain.  The first excited states at an energy 
$K_{\parallel}$ are located on the sites $n = \pm 1$ and corresponding 
to the odd site chain.  Level crossings occur whenever
\begin{equation}
h = (2n+1) K_{\parallel}; \ \ \ \ n=0,1,2,3 \ldots,
\label{deux}
\end{equation}
which has the period $2 K_{\parallel}$.  With this period the 
splitting between the ground state and the first excited state 
undergoes sawtooth oscillations with amplitude $\sim K_{\parallel}$ 
and with each level crossing the ground state passes from one chain to 
the other.  Given that as a function of $K_{\perp}$ there are no 
singularities in the characteristic determinant, this fixed point will 
govern the behavior for all values of this parameter.  That this is 
the case for a ferromagnet is verified by both the analytic and 
numerical calculations described below.

The vicinity of the fixed point with
\begin{equation}
S^{2} K_{\perp} < K_{\parallel}
\label{trois}
\end{equation}
defines the {\it small particle limit\/}.  The corrections to the 
ground state $|n=0>$ are perturbative, however the excited states 
split.  These are $(1/\sqrt{2})[|1> \pm |-1>]$ with energies 
$K_{\parallel} \pm K_{\perp}$.  The smallest $h=0$ splitting 
$K_{\parallel} - |K_{\perp}|$ determines the amplitude of the near
saw-tooth oscillations.  This splitting has nothing to do with 
tunneling.  However the ground state is clearly ``coherent'' in the 
sense $<\vec S>=0$, i.e., the order parameter is not localized in a 
particular direction in the $x-y$-plane.  And if, e.g., the initial 
state is  $(1/\sqrt{3})[ |0> + |1> + |-1>]$, $<\vec S>$ has a 
magnitude of $\sim S$ and points in the positive $x$-direction.  With 
a frequency $\omega_1 = H_{\parallel}$, it oscillates rapidly between 
the $+x$ and $-x$-directions.  While with the frequency $\omega_2 = 
H_{\perp}$ these oscillations precess in the perpendicular plane so 
that one quarter of a long period later the rapid oscillations are 
between the $\pm y$-directions.  Such oscillatory behavior is 
characteristic of coherence.

Quite generally Schr\"odinger's equation
\begin{eqnarray}
 &(\epsilon - ( K_{\parallel}n^{2} - nh)) a_{n} \nonumber\\
& = {1\over4} 
K_{\perp} M_{n}^{n+1} M_{n+1}^{n+2} [(a_{n+2}- a_{n})+ 
(a_{n-2}-a_{n})], \label{quatre} 
\end{eqnarray}
involves finite differences, where the wave-function $\Psi = \sum_{n} 
(-)^{n}a_{n}f^{\dagger}_{n}|>$.  When the inequality (\ref{trois}) is 
reversed the particle is {\it large} and a continuum approximation to 
this Schr\"odinger's equation is appropriate.  Within the definition 
of a Fourier transform, $a_{n} = (1/ \sqrt{2\pi}) \int_{-\infty 
}^{-\infty }dp f(p) e^{-i p (n-d)}$, the function $f(p)$ is defined 
for the even site chain and $h=0$.  For finite $h$ the center of the 
real space potential is displaced by $d = h/2 K_{\parallel}$ and the 
odd site chain is accommodated via a $d=1 + (h/2 K_{\parallel})$.  
Assuming $ K_{\perp}<0$, the continuum approximation to Eqn.  
(\ref{quatre}) is:
\begin{equation}
(\epsilon - {h^{2}\over 4 K_{\parallel}}
+ K_{\parallel}{\partial^{2} \over \partial p^{2}}) f(p) = - 
{S^{2}K_{\perp}\over 2} [\cos 2 p - 1] f(p),
\label{cinq}
\end{equation}
which is Mathieu's equation\cite{AS}.  The potential is periodic with 
period $\Delta p = \pi$ and the low lying solutions form bands.  For a 
given energy, a solution might be characterized by a wave-vector $k$ 
and Floquet's (Bloch's) theorem\cite{AS} implies that solutions are of 
the form $f_{k}(p) = e^{ikp}u_{k}(p)$ where $u_{k}(p) = u_{k}(p+ 
\pi)$.  That $a_{n}$ be finite implies $k = d$ (to within trivial 
translations).  Within a given well of the periodic potential, the low 
lying states have wave-functions which to a good approximation are those 
of a harmonic oscillator.  Taking $K_{\perp}<0$, around $p=0$, $ f(p) 
= (1/ \sqrt{\beta \sqrt{\pi}}) e^{-p^{2}/2\beta^{2}}$ with $\beta^{2} 
= S \sqrt{{ K_{\parallel} / K_{\perp}}}$, and the nominal ground state 
energy is $(h^{2}/ 4K_{\parallel}) +(\omega_{0}/2)$; $\omega_{0} = 2S 
\sqrt{K_{\parallel} K_{\perp}}$.  Because of tunneling between wells 
this ground state level becomes a band, i.e., the energies 
$\epsilon_{k} = (h^{2}/ 4 K_{\parallel}) + (\omega_{0}/2) + (w/2)\cos 
\pi k$ where the width\cite{AS} $w = 8 \sqrt{2/\pi} 
\omega_{0}S_{f}^{1/2}e^{-S_{f}}$ and where the action $S_{f} = 2 
S\sqrt{(K_{\perp}/K_{\parallel})}$.

\begin{figure}[t]
\epsfig{width=0.95 \linewidth,file=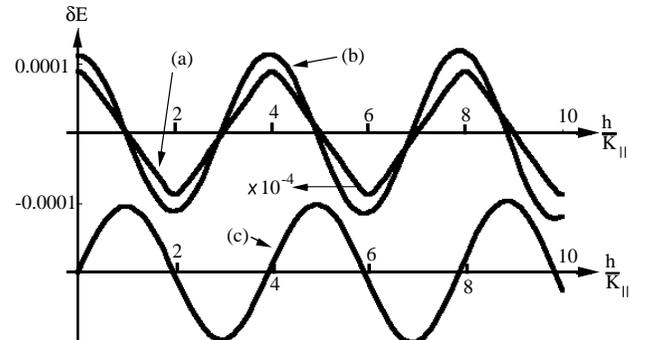} \caption{ Each plot 
corresponds to $K_{\parallel}=1$.  For (a) $K_{\perp} = .0003$, 
$S=40$, and the saw-tooth tunnel spilting $\sim K_{\parallel}$ has 
been divided by $10^{4}$.  Both (b), with integer spin $S=40$, and
and (c), with the half-integer value $S=81/2$, have $K_{\perp} = .03$ and 
the form is sinusoidal.   Notice the $\pi/2$ phase shift.  }
\label{C}
\end{figure}

The result for the ``tunnel splitting'', i.e., the difference in 
energy between the ground states for the even and odd chains is
\begin{eqnarray}
\delta E &=& 4 \sqrt{2 \over \pi} \omega_{0}S_{f}^{1/2}e^{-S_{f}} 
\cos ( \pi {S h \over 2 H_{\parallel}}) ;\nonumber\\
 S_{f} &=&  
2 S\sqrt{(H_{\perp}/ H_{\parallel})}{.}
\label{huit}
\end{eqnarray}
The numerical work shown in Fig .  \ref{C} confirms this result\break 

\vskip -15pt

\noindent that large particles have a tunnel splitting which 
oscillates as $\cos ( S\pi h / 2 H_{\parallel}) $.  This with the 
equivalents for an anti-ferromagnet, are the principal results of this 
Letter.

A perpendicular magnetic field, i.e., one lying in the $x-y$-plane, 
couples the two chains.  Still, for a larger particle and 
$K_{\perp}<0$, the wave-function, $a_{n}$, is extended, does not 
alternate in sign between sites, and is well localized near the center 
of a chain.  The operators $S^{\pm}$, to a good approximation, cause 
translations by one lattice site, i.e., they convert the ground state 
on one chain into that on the other.  The matrix elements of 
$h_{x}S_{x} =(h_{x}/2)[S^{+} + S^{-}]$ are approximately $h_{x}S/2$ 
while those of $h_{y}S_{y}$ are negligible.  Assuming that $h$ is 
smaller than the splitting $\sim \omega_{0}$ between the the ground 
doublet and other excited states, the splitting of the doublet is:
\begin{equation}
\delta E_{h} = \sqrt{(\delta E)^{2} + (S h_{x})^{2}}.
\label{neuf}
\end{equation}
Evidently, the field scale it determined by the tunnel splitting 
$\delta E/S $ and implies a rather careful alignment is required in 
order to observe oscillations.  If $K_{\perp}>0$, $a_{n}$ alternates 
in sign between sites and the role of the $x$ and $y$ axes are 
interchanged.

Again for a large particle (and $K_{\perp}<0$), since $S_{x}$ has 
large, and $S_{y}$ very small, matrix elements within the low lying 
doublet, it follows that if the initial condition puts the system in 
an approximately equally weighted linear combination of these two 
states, the magnetization will oscillate in the $x$-direction at a 
frequency corresponding to the tunnel splitting, {\it but\/} the 
component in the $y$-direction is negligible.  Thus, in passing from a 
small to a large particle the magnetization becomes localized in the 
$x$-direction.  Again for $K_{\perp}>0$ the role of the $x$ and $y$ 
axes are interchanged.

Turning to spin-parity effects, consider a {\it small particle with 
half-integer spin}.  The fixed point now has level crossings occurring 
whenever $h=2n K_\parallel$ $n=0,1,2,3 \ldots$ which reflects a shift 
of $\pi/2$ relative to the integer spin case.  Kramers' degeneracy 
corresponds to the level crossing at $h=0$, but apart from the $\pi/2$ 
phase shift a half integer small particle behaves like its integer 
spin equivalent.

Similarly for a {\it large particle and half-integer spin\/}, all is 
the same as for the integer spin case except for the $\pi/2$ phase 
shift.  In particular in the tunnel splitting, Eqn.  (\ref {huit}), 
the cosine is replaced by a sine.

An {\it easy-axis\/} ferromagnet\cite{chud} is quite different from 
its {\it easy-plane\/} equivalent.  In particular there is no 
oscillatory field dependence.

For larger fields an {\it easy-plane\/} anti-ferromagnet is much like 
a similar ferromagnet and there is again a quasi-periodic modulation 
of the tunnel splitting.  The effective Hamiltonian\cite{afm} ${\cal 
H} =g \mu_{B}\vec S\cdot \vec H + J\vec S_A \cdot \vec S_B + 
K_\parallel \left[\left({S_{A,z}}\right)^2 + 
\left({S_{B,z}}\right)^2\right]$ contains the external field, exchange 
and a suitable anisotropy energy.  The exchange field $J$ couples the 
sub-lattices ``$A$'' and ``$B$'' and it is assumed that $S_{A} = 
S_{B}=S$ unless stated otherwise.  The auxiliary particles $ 
a^{}_{n,m} $ create single particle states $|n,m>\equiv 
a^\dagger_{n,m}|>$ which map to the $|S^{A}_{z}= n+m,S^{B}_{z}= -n >$.  
The constraint is $ Q_m = \sum_n a^{\dagger}_{n,m} a^{}_{n,m} = 1$, 
and the auxiliary particle Hamiltonian\cite{sebetal},
\begin{equation}
{\cal H}_m = \sum_n \left[ \epsilon_{n} a^{\dagger}_{n,m} a^{}_{n,m} + 
t_{n}^{n+1} ( a^{\dagger}_{n+1,m} a^{}_{n,m} + H.c.)  \right]
\label{dix}
\end{equation}
where, the diagonal energies are $ \epsilon_{n} =-J(n+m)n - 
K_\parallel [(n+m)^2 + n^2]$, and the hopping matrix elements $ 
t_{n}^{n+1} = (J / 2) M_{n+m}^{n+m+1} M_{n}^{n+1} $.  Hopping couples 
{\it neighboring\/} sites, {\it and\/} there is a distinct chain for 
every value of the quantum number $m \equiv S_{z}$.

The {\it chains\/} are roughly equivalent to the {\it sites\/} for the 
ferromagnet.  The chains with $m$ values which differ by 2 are coupled 
by the inter-chain hopping terms which arises when the term 
$K_{\perp}[{S_{A,x}}^{2} + {S_{B,x}}^{2}]$ is added.  The structure 
comprises two distinct two dimensional networks, the equivalent of the 
chains for a ferromagnet, and as the magnetic field changes the ground 
state can pass form one network to the other causing the tunnel 
splitting to oscillate in a quasi-periodic manner.  Here it is 
{\it not\/} assumed that $K_{\perp}<K_\parallel$.  The larger and 
smaller of the two will be denoted $K$ and $k$ respectively.

The period of the oscillations is determined by the condition for 
level crossings.  The details are different for small and large 
particles.  A {\it small particle\/} is defined by
\begin{equation}
J > S^{2}K_\parallel
\label{onze}
\end{equation}
For a given chain, with index $m$ ($\equiv S_{z}$), the exchange 
energy $J\vec S_A \cdot \vec S_B$ dominates the anisotropy energy 
$K_\parallel$ and trivially the ground state has an 
energy\cite{sebetal} $(J/2) |m|(|m|+1) - mh$.  The fixed point level 
crossings occur when
\begin{equation}
h = J(n+1); \ \ \ \ n = 1,2,3\ldots,
\label{douze}
\end{equation}
i.e., the first {\it possible\/} crossing, and a zero of the tunnel 
splitting, occurs when $h=2J$ and otherwise has period $\Delta h = J$.  
Notice there is a phase shift of $\pi/2$ relative to level crossings 
for a similar ferromagnet, Eqn(\ref{deux}), and that $J/2$ has replaced 
the anisotropy parameter $K_\parallel$.  The magnitude of the tunnel 
splitting depends upon another small to large cross-over.  If 
$S^{2}K_{\perp} < J$ the particle is {\it twice times small\/} and the 
periodic saw-tooth like splitting is $\sim J=H_{e}/S$. Here the 
anisotropy energy is irrelevant and similar periodic behavior will 
occur for a twice small {\it easy axis\/} anti-ferromagnet.

When this latter inequality is reversed the particle is {\it 
small-large\/} and it is implied that $K_{\perp} = K = H_{a}/S$, $ 
K_\parallel = k$.  There are two regimes.  For the {\it small field 
regime\/}, $h < \omega_{0} = 2S \sqrt{KJ}= 2\sqrt{H_{a}H_{e}}$, the 
axis of quantization is rotated to lie along the $x$-direction, which 
interchanges the role of, $K_{\perp} = k$ ($ = 0$ for simplicity) and 
$K_\parallel = K$ and the field term is $hS_{x}$.  As a result, the 
inverse of inequality (\ref{onze}) is satisfied, the wave-functions on 
the chains are those of an harmonic oscillator, and the chain ground 
state energies are\cite{notseb} $(\omega_{0}/2) + J m^{2}/2$.  The 
{\it very small field\/} limit applies when $h<J = H_{e}/S$ whence the 
coupling between the chains is a perturbation.  The field couples each 
chain $m$ to the {\it first excited state\/} on two adjacent chains 
$m\pm1$, with a matrix element $h\beta / 2\sqrt{2} $, $\beta^{2} = 
2S\sqrt{K/J}$.  The diagonal energy correction is $-2(h\beta / 
2\sqrt{2} )^{2}/ \omega_{0}$.  An off-diagonal term $t_{h}= - (h\beta 
/ 2\sqrt{2} )^{2}/ \omega_{0}$ couples the $m = \pm 1$ chains via an 
excited state on the $m=0$ chain.  This splits the pair, with the 
lower energy being $(J/2) - 3 ( h\beta )^{2}/ 8\omega_{0}$.  For very 
small fields the splitting is therefore
\begin{equation}
\delta E =
{J \over 2} - {(h\beta )^{2}\over 8\omega_{0}} = {J \over 2} - {1\over 
8} {h^{2}\over J} ={SH_{e} \over 2} - {1\over 8} {S h^{2}\over H_{e}}.
\label{treize}
\end{equation}

For {\it larger fields}, but still in the small field regime, the 
off-diagonal coupling $t_{h}$ exceeds the difference in energy between 
adjacent sites and a continuum approximation is appropriate.  The 
problem reduces\cite{Chiolero} to that for a ferromagnet and the 
result is Eqn.  (\ref{huit}) with $h=0$, $H_\parallel \to H_{e}/2$ and 
$H_{\perp} \to - h^{2}/2 H_{e}$.  The zero point energy for 
motion perpendicular to the chains is $\sim h$ while that associated 
with motion along the chains is $\omega_{0}$.  The equality of these 
two energies indicates the point at which the appropriate axis of 
quantization changes and implies the small to large field inequality 
quoted above.

In the large field regime, $h > \omega_{0}$, reverting to an axis of 
quantization defined by the field direction, the ground state energy 
for the chain labeled $m$ is again $(J/2) |m|(|m|+1) - mh = 
(J/2)[m-(h/J) + (1/2)]^{2} -(h^{2}/2J) - (J/8)$ and represents an 
harmonic potential, centered at $(h/J) - (1/2)$, for the motion 
perpendicular to the chains.  The coupling between chains is $S^{2} 
K_{\perp}/2$ and Eqn.  (\ref{huit}) again applies with $H_{\parallel} 
\to H_{e}/2$, $H_{\perp} \to 2 H_{\perp}$, and with the cosine 
replaced by a sine\cite{Chiolero}.  The same result is valid for small 
{\it integer\/} values of $j=|S_{A}- S_{B}|$ the excess, or net, spin.  
For small {\it half-integer\/} $j$ values after the substitution 
$H_{\parallel} \to H_{e}/2$, Eqn.  (\ref{huit}) is valid, as written.

For a {\it larger\/} magnet when inequality (\ref{onze}) is reversed, 
the chain ground state energy is again $(\omega_{0}/2) + 
(1/2)m^{2}J-mh$, identical, apart from a constant, to the site energy 
for the ferromagnetic problem with the replacement $K_{\parallel} \to 
J/2$.  The level crossings correspond to $h=J(2n+1)/2$, i.e., have 
period $J$.  Also as for the previous cases, the magnitude of the 
tunnel splitting depends upon another small to large cross-over.  Now 
if $S^{2}K_{\perp} < J$ the particle is {\it large-small} and the 
periodic splitting $\sim J = H_{e} /S$.  When this inequality is 
reversed, corresponding to the {\it large-large\/} case, after the 
substitutions $H_{\parallel} \to H_{e}/2$ and $H_{\perp} \to 2 
H_{\perp}$, the result is Eqn.  (\ref{huit}) for integer $j$ while the 
cosine is replaced by a sine for half-integer $j$.

Suitable single crystals of molecular magnets would ideal for studying 
the predictions made here.  The only relevant experimental evidence 
known to the author is for the anti-ferromagnet Fe${}_{10}$\cite{Taft} 
and would appear to belong to the small-small limit.  The series of 
small ferromagnets based on Mn, all seem to have an easy 
axis.  However many other systems exist for which the nature of the 
anisotropy have never been determined.



\begin{references}
\bibitem{notseb} Some more details of the calculations can be deduced 
from material to be found in ref.  4.
\end{references}
\end{document}